\documentclass[letterpaper]{aipproc}

\layoutstyle{6x9}

\SetInternalRegister\hbadness{8000} 

\newcommand\doingARLO[2][]{%
  \ifx\mmref\undefined #1\else #2\fi
}

\begin{document}

\title 
      [Finite-N Conformality and Gauge Coupling Unification]
      {Finite-N Conformality and Gauge Coupling Unification}


\author{Paul H. Frampton}{
  address={Department of Physics and Astronomy, CB \#3255,
University of North Carolina, Chapel Hill, NC 27599-3255, USA.}
}

\iftrue

\fi

\copyrightyear  {2001}

\begin{abstract}
In this talk I review some aspects of the idea that there is an infra-red
conformal fixed-point at the TeV scale. In particular, it is shown how
gauge coupling unification can be achieved by TeV unification
in a semi-simple gauge group.
\end{abstract}

\date{\today}

\maketitle

\section{Introduction}

\bigskip

Japan is the ideal place to talk about string theory since much of it originated here;
for example, as a postdoc in Chicago my mentor Yoichiro Nambu explained to me in 1969(!)
the relevance of two-dimensional conformal invariance.

In particle phenomenology, the impressive success of the standard theory
based on $SU(3) \times SU(2) \times U(1)$ has naturally led to the question
of how to extend the theory to higher energies? One is necessarily led by
weaknesses and incompleteness in the standard theory. If one extrapolates
the standard theory as it stands one finds (approximate) unification of the
gauge couplings at $\sim 10^{16}$ GeV. But then there is the {\it hierarchy}
problem of how to explain the occurrence of the tiny dimensionless ratio $%
\sim 10^{-14}$ of the weak scale to the unification scale. Inclusion of
gravity leads to a {\it super-hierarchy} problem of the ratio of the weak
scale to the Planck scale, $\sim 10^{18}$ GeV, an even tinier $\sim 10^{-16}$
Although this is obviously a very important problem about which
conformality by itself is not informative, we shall discuss first the
hierarchy rather than the super-hierarchy.

\bigskip

There are four well-defined approaches to the hierarchy problem:

\begin{itemize}
\item  1. Supersymmetry

\item  2. Technicolor.

\item  3. Extra dimensions.

\item  4. Conformality.
\end{itemize}

\noindent {\it Supersymmetry} has the advantage of rendering the hierarchy
technically natural, that once the hierarchy is put in to the lagrangian it
need not be retuned in perturbation theory. Supersymmetry predicts
superpartners of all the known particles and these are predicted to be at or
below a TeV scale if supersymmetry is related to the electroweak breaking.
Inclusion of such hypothetical states improves the gauge coupling
unification. On the negative side, supersymmetry does not explain the origin
of the hierarchy.

\bigskip

\noindent {\it Technicolor} postulates that the Higgs boson is a composite
of fermion-antifermion bound by a new (technicolor) strong dynamics at or
below the TeV scale. This obviates the hierarchy problem. On the minus side,
no convincing simple model of technicolor has been found.

\bigskip

\noindent {\it Extra dimensions} can have a range as large as $1 ({\rm TeV}%
)^{-1}$ and the gauge coupling unification can happen quite differently than
in only four spacetime dimensions. This replaces the hierarchy problem with
a different fine-tuning question of why the extra dimension is restricted to
a distance corresponding to the weak interaction scale.

\bigskip

\noindent {\it Conformality} is inspired by superstring duality and assumes
that the particle spectrum of the standard model is enriched such that there
is a conformal fixed point of the renormalization group at the TeV scale.
Above this scale the coupling do not run so the hierarchy is nullified.

\bigskip

Conformality is the approach followed in this paper. We shall systematicaly
analyse the compactification of the IIB superstring on $AdS_5 \times
S^5/\Gamma$ where $\Gamma$ is a discrete non-abelian group.

The duality between weak and strong coupling field theories and then between
all the different superstring theories has led to a revolution in our
understanding of strings. Equally profound, is the AdS/CFT duality which is
the subject of the present article. This AdS/CFT duality is between string
theory compactified on Anti-de-Sitter space and Conformal Field Theory.

Until very recently, the possibility of testing string theory seemed at best
remote. The advent of $AdS/CFT$s and large-scale string compactification
suggest this point of view may be too pessimistic, since both could lead to $%
\sim 100TeV$ evidence for strings. With this thought in mind, we are
encouraged to build $AdS/CFT$ models with realistic fermionic structure, and
reduce to the standard model below $\sim 1TeV$.

Using AdS/CFT duality, one arrives at a class of gauge field theories of
special recent interest. The simplest compactification of a ten-dimensional
superstring on a product of an AdS space with a five-dimensional spherical
manifold leads to an ${\cal N} = 4~SU(N)$ supersymmetric gauge theory, well
known to be conformally invariant\cite{mandelstam}. By replacing the
manifold $S^5$ by an orbifold $S^5/\Gamma$ one arrives at less
supersymmetries corresponding to ${\cal N} = 2,~1 ~{\rm or}~ 0$ depending
\cite{KS} on whether $\Gamma \subset SU(2), ~~ SU(3), ~~{\rm or} 
\not{\subset }SU(3)$ respectively, where $\Gamma$ is in all cases a subgroup of 
$SU(4) \sim SO(6)$ the isometry of the $S^5$ manifold.

It was conjectured in \cite{maldacena} that such $SU(N)$ gauge theories are
conformal in the $N \rightarrow \infty$ limit. In \cite{F1} it was
conjectured that at least a subset of the resultant nonsupersymmetric ${\cal %
N} = 0$ theories are conformal even for finite $N$. Some first steps to
check this idea were made in \cite{WS}. Model-building based on abelian $%
\Gamma$ was studied further in \cite{CV,F2,F3}, arriving in \cite{F3} at an $%
SU(3)^7$ model based on $\Gamma = Z_7$ which has three families of chiral
fermions, a correct value for ${\rm sin}^2 \theta$ and a conformal scale $%
\sim 10$~~TeV.

The case of non-abelian orbifolds bases on non-abelian $\Gamma$ has not
previously been studied, partially due to the fact that it is apparently
somewhat more mathematically sophisticated. However, we shall show here that
it can be handled equally as systematically as the abelian case and leads to
richer structures and interesting results.

In such constructions, the cancellation of chiral anomalies in the
four-dimensional theory, as is necessary in extension
of the standard model ({\it e.g.} \cite{chiral,331}),
follows from the fact that the progenitor ten-dimensional
superstring theory has cancelling hexagon anomaly\cite{hexagon}.
It offers a novel approach to family unification\cite{guts,Pak}.

We consider all non-abelian discrete groups of order $g < 32$. These are
described in detail in \cite{books,FK}. There are exactly 45 such
non-abelian groups. Because the gauge group arrived at by this 
construction\cite{CV} is $\otimes_i SU(Nd_i)$ where $d_i$ are the dimensions of the
irreducible representations of $\Gamma$, one can expect to arrive at models
such as the Pati-Salam $SU(4) \times SU(2) \times SU(2)$ type\cite{PS} by
choosing $N = 2$ and combining two singlets and a doublet in the {\bf 4} of 
$SU(4)$. Indeed we shall show that such an accommodation of the standard
model is possible by using a non-abelian $\Gamma$.

The procedures for building a model within such a conformality approach are:
(1) Choose $\Gamma$; (2) Choose a proper embedding $\Gamma \subset SU(4)$ by
assigning the components of the {\bf 4} of $SU(4)$ to irreps of $\Gamma$,
while at the same time ensuring that the {\bf 6} of $SU(4)$ is real; (3)
Choose $N$, in the gauge group $\otimes_i SU(Nd_i)$. (4) Analyse the
patterns of spontaneous symmetry breaking.

In the present study we shall choose $N = 2$ and aim at the gauge group 
$SU(4) \times SU(2) \times SU(2)$. To obtain chiral fermions, it is 
necessary\cite{CV} that the {\bf 4} of $SU(4)$ be complex ${\bf 4} \neq {\bf 4}^*$.
Actually this condition is not quite sufficient to ensure chirality in the
present case because of the pseudoreality of $SU(2)$. We must ensure that
the {\bf 4} is not just pseudoreal.

This last condition means that many of our 45 candidates for $\Gamma$ do not
lead to chiral fermions. For example, $\Gamma = Q_{2n} \subset SU(2)$ has
irreps of appropriate dimensionalities for our purpose but it will not
sustain chiral fermions under $SU(4)\times SU(2) \times SU(2)$ because these
irreps are all, like $SU(2)$, pseudoreal.\footnote{%
Note that were we using $N \geq 3$ then a pseudoreal {\bf 4} would give
chiral fermions.} Applying the rule that {\bf 4} must be neither real nor
pseudoreal leaves a total of only 19 possible non-abelian discrete groups of
order $g \leq 31$. The smallest group which avoids pseudoreality has order 
$g = 16$ but gives only two families. The technical details of our systematic
search will be postponed to a future publication. Here we shall present only
the simplest interesting non-abelian case which has $g = 24$ and gives three
chiral families in a Pati-Salam-type model\cite{PS}.

Before proceeding to the details of the specific $g = 24$ case, it is worth
noting that the CFT it
exemplifies should be free of all divergences if the
conformality conjecture is correct and be UV finite. Further the
theory is originating from a superstring theory in a higher-dimension (ten)
and contains gravity\cite{V,RS,GW} by compactification of the
higher-dimensional graviton already contained in that superstring theory. In
the CFT as we derive it in
d=4 flat spacetime, gravity is absent because we have not kept these
graviton modes - of course, their influence on high-energy physics
experiments is generally completely negligible unless the compactification
scale is ``large''\cite{antoniadis}.

To motivate our model it is instructive to comment on the choice of $\Gamma$
and on the choice of embedding.

If we embed only four singlets of $\Gamma$ in the {\bf 4} of $SU(4)$ then
this has the effect of abelianizing $\Gamma$ and the gauge group obtained in
the chiral sector of the theory is $SU(N)^q$. These cases can be interesting
but have already been studied\cite{CV,F2}. Thus, we require at least one
irrep of $\Gamma$ to have $d_i \geq 2$ in the embedding.

The only $\Gamma$ of order $g \leq 31$ with a {\bf 4} is $Z_5 \tilde{\times}
Z_4$ and this embedding leads to a non-chiral theory. This leaves only
embeddings with two singlets and a doublet, a triplet and a singlet or two
doublets.

The third of these choices leads to richer structures for low order $\Gamma$. 
Concentrating on them shows that of the chiral models possible, those from
groups of low order result in an insufficient number (below three) of chiral
families.

The first group that can lead to exactly three families occurs at order $g =
24$ and is $\Gamma = Z_3 \times Q$ where $Q (\equiv Q_4)$ is the group of
unit quarternions which is the smallest dicyclic group $Q_{2n}$.

There are several potential models due to the different choices for the 
{\bf 4} of $SU(4)$ but only the case {\bf 4} = $(1\alpha, 1^{^{\prime}}, 2\alpha)$
leads to three families so let us describe this in some detail:

Since $Q \times Z_3$ is a direct product group, we can write the irreps as 
$R_i \otimes \alpha^{a}$ where $R_i$ is a $Q$ irrep and $\alpha^{a}$ is a 
$Z_3 $ irrep. We write $Q$ irreps as $1,~1^{^{\prime}},~1^{^{\prime\prime}},~
1^{^{\prime\prime\prime}},~2$ while the irreps of $Z_3$ are all singlets
which we call $\alpha, \alpha^2, \alpha^3 = 1$. Thus $Q \times Z_3$ has
fiveteen irreps in all and the gauge group will be of Pati-Salam type for $N
= 2$.

If we wish to break all supersymmetry, the {\bf 4} may not contain the
trivial singlet of $\Gamma$. Due to permutational symmetry among the
singlets it is sufficiently general to choose {\bf 4} = $(1%
\alpha^{a_1},~1^{^{\prime}}\alpha^{a_2},~2\alpha^{a_3})$ with $a_1 \neq 0$.

To fix the $a_i$ we note that the scalar sector of the theory which is
generated by the {\bf 6} of $SU(4)$ can be used as a constraint since the
{\bf 6} is required to be real. This leads to 
$a_1 + a_2 = - 2a_3 ({\rm mod} ~3)$. 
Up to permutations in the chiral fermion sector the most general
choice is $a_1 = a_3= +1$ and $a_2 = 0$. Hence our choice of embedding is
\begin{equation}
{\bf 4} = (1\alpha,~1^{^{\prime}},~2\alpha)  \label{embed}
\end{equation}
with
\begin{equation}
{\bf 6} =
(1^{^{\prime}}\alpha,~2\alpha,~2\alpha^{2},~1^{^{\prime}}\alpha^{2})
\label{six}
\end{equation}
which is real as required.

We are now in a position to summarize the particle content of the theory.
The fermions are given by
\begin{equation}
\sum_I~{\bf 4}\times R_I
\end{equation}
where the $R_I$ are all the irreps of $\Gamma = Q \times Z_3$. This is:
\[
\sum_{i=1}^{3} [(2_{1}\alpha^{i},2_{2}\alpha^{i})
+(2_{3}\alpha^{i},2_{4}\alpha^{i})+(2_{2}\alpha^{i},2_{1}\alpha^{i})
+(2_{4}\alpha^{i},2_{3}\alpha^{i})+(4\alpha^{i},\overline{4}\alpha^{i})]
\]

\begin{equation}
+ \sum_{i=1}^{3} \sum_{a=1}^{4} [(2_{a}\alpha^{i},
2_{a}\alpha^{i+1})+(2_{a}\alpha^{i},4\alpha^{i+1}) + (\bar{4}%
\alpha^{i},2_{a}\alpha^{i+1})]  \label{fermions}
\end{equation}

It is convenient to represent the chiral portions of these in a given
diagram (see Figure 1).

The scalars are given by
\begin{equation}
\sum_I~{\bf 6}\times R_I
\end{equation}
and are:
\[
\sum_{i=1}^{3} \sum_{j=1(j\neq i)}^{3} [(2_{1}\alpha^{i},2_{2}
\alpha^{j})+(2_{2}\alpha^{i}, 2_{1}\alpha^{j})+(2_{3}\alpha^{i},
2_{4}\alpha^{j})+(2_{4}\alpha^{i},2_{3}\alpha^{j})
+(2_{2}\alpha^{i},2_{1}\alpha^{i})+(2_{4}\alpha^{i},2_{3}\alpha^{i})]
\]
\begin{equation}
+ \sum_{i=1}^{3} \sum_{j=1(j\neq i)}^{3} \{
\sum_{a=1}^{4}[(2_{a}\alpha^{i},4\alpha^{j}) +\bar{(4}\alpha^{i},2_{a}
\alpha^{j} )] +(4\alpha^{i}, \bar{4}\alpha ^{i}) \}  \label{scalars}
\end{equation}
which is easily checked to be real.

The gauge group $SU(4)^3 \times SU(2)^{12}$ with chiral fermions of Eq.(\ref
{fermions}) and scalars of Eq.(\ref{scalars}) is expected to acquire
confromal invariance at an infra-red fixed point of the renormalization
group, as discussed in \cite{F1}.

To begin our examination of the symmetry breaking we first observe that if
we break the three $SU(4)$s to the totally diagonal $SU(4)$, then chirality
in the fermionic sector is lost. To avoid this we break $SU_{1}(4)$
completely and then break $SU_{\alpha }(4)\times SU_{\alpha ^{2}}(4)$ to its
diagonal subgroup $SU_{D}(4).$ The first of these steps can be achieved with
VEVs of the form $[(4_{1},2_{b}\alpha ^{k})+h.c.]$ where we are free to
choose $b$, but $k$ must be $1$ or $2$ since there are no $%
(4_{1},2_{b}\alpha ^{k=0})$ scalars. The second step requires an

$SU_{D}(4)$ singlet VEV from ($\overline{4}_{\alpha }$,4$_{\alpha^{2}})$
and/or (4$_{\alpha }$, $\overline{4}_{\alpha ^{2}})$. Once we make a choice
for $b$ (we take $b=4$), the remaining chiral fermions are, in an intuitive
notation:

\bigskip

\noindent $\ \sum_{a=1}^{3}\left[ (2_{a}\alpha \ ,1,4_{D})+(1,2_{a}\alpha
^{-1},\overline{4_{D}})\right] $

\bigskip

\noindent which has the same content as as a three family Pati-Salam model,
though with a separate $SU_{L}(2)\times SU_{R}(2)$ per family.

To further reduce the symmetry we must arrange to break to a single 
$SU_{L}(2)$ and a single $SU_{R}(2).$ This is achieved by modifying step one
where $SU_{1}(4)$ was broken. Consider the block diagonal decomposition of 
$SU_{1}(4)$ into $SU_{1L}(2) \times SU_{1R}(2).$ The representations 
$(2_{a}\alpha ,4_{1})$ and $(2_{a}\alpha ^{-1},4_{1})$ then decompose as
$(2_{a}\alpha ,4_{1})\rightarrow (2_{a}\alpha ,2,1)+(2_{a}\alpha ,1,2)$ and 
$(2_{a}\alpha ^{-1},4_{1})\rightarrow (2_{a}\alpha ^{-1},,2,1)+(2_{a}\alpha
^{-1},1,2)$. Now if we give $VEVs$ of equal magnitude to the $(2_{a}\alpha
,,2,1),$ $a=1,2,3$, and equal magnitude $VEVs$ to the $(2_{a}\alpha
^{-1},1,2)$ $a=1,2,3,$ we break $SU_{1L}(2) \times
\prod\limits_{a=1}^{3}SU(2_{a}\alpha )$ to a single $SU_{L}(2)$ and we break
$SU_{1R}(2) \times \prod\limits_{a=1}^{3}SU(2_{a}\alpha )$ to a single 
$SU_{R}(2).$ Finally, $VEVs$ for $(2_{4}\alpha ,2,1)\ $and $(2_{4}\alpha
,1,2) $ as well as $(2_{4}\alpha ^{-1},2,1)\ $and $(2_{4}\alpha ^{-1},1,2)$
insures that both $SU(2_{4}\alpha )$ and $SU(2_{4}\alpha ^{-1})$ are broken
and that only three families remain chiral. The final set of chiral fermions
is then $3[(2,1,4)+(1,2,\bar{4})]$ with gauge symmetry $SU_{L}(2) \times
SU_{R}(2) \times SU_{D}(4).$

To achieve the final reduction to the standard model, an adjoint VEV\ from 
($\overline{4}_{\alpha }$,4$_{\alpha ^{2}})$ and/or 
(4$_{\alpha }$,$\overline{4}_{\alpha ^{2}})$ 
is used to break $SU_{D}(4)$ to the $SU(3)\times U(1),$
and a right handed doublet is used to break $SU_{R}(2).$

While this completes our analysis of symmetry breaking, it is worthwhile
noting the degree of constraint imposed on the symmetry and particle content
of a model as the number of irreps $N_{R}$ of the discrete group $\Gamma $
associated with the choice of orbifold changes. The number of guage groups
grows linearly in $N_{R}$, the number of scalar irreps grows roughly
quadratically with $N_{R}$, and the chiral fermion content is highly $\Gamma
$ dependent. If we require the minimal $\Gamma $ that is large enough for
the model generated to contain the fermions of the standard model and have
sufficient scalars to break the symmetry to that of the standard model, then
$\Gamma = Q \times Z_{3}$ appears to be that minimal choice\cite{FK2}.

Although a decade ago the chances of testing string theory seemed at best
remote, recent progress has given us hope that such tests may indeed be
possible in AdS/CFTs. The model provided here demonstrates the standard
model can be accomodated in these theories and suggests the possibility of a
rich spectrum of new physics just around the TeV corner.

\bigskip

\section{Gauge Coupling Unification}

\bigskip

There is not space here to describe many technical details which are,
however, available
in the published papers cited at the end of this talk. But I would like to emphasize one
success of the approach which involves the unification of gauge couplings
\cite{F2,FMS}. Recall that the successful such unification is one primary
reason for belief in supersymmetric
grand unification {\it e.g.} \cite{Wil}. That argument is simple to state:
The RG equations are:

\begin{equation}
\frac{1}{\alpha_i(M_G)} = \frac{1}{\alpha_i(M_Z)} - \frac{b_i}{2 \pi} \ln 
\left( \frac{M_G}{M_Z} \right)
\end{equation}
Using the LEP values at the Z-pole as $\alpha_3 = 0.118 \pm 0.003$,
$\alpha_2 = 0.0338$ and $\alpha_1 = \frac{5}{3}\alpha_Y = 0.0169$
(where the eroors on $\alpha_{1,2}$ are less than 1\%) and
the MSSM values $b_i = (6\frac{3}{5}, 1, -3)$ leades
to $M_G = 2.4 \times 10^{16}$ GeV and the {\it predictiion} that
$\sin^2\theta = 0.231$
in excellent agreement with experiment.

In the present approach the three gauge couplings $\alpha_{1,2,3}$
run up to $\sim 1$TeV where they freeze and embed in a larger (semi-simple)
gauge group which contains $SU(3) \times SU(2) \times U(1)$.

I will give two examples, the first based on the abelian orbifold $S^5/Z_7$
and the second based on the non-abelian orbifold $S^5/(D_4 \times Z_3)$.

In the first, abelian, case we choose N=3, $\Gamma=Z_7$ 
and the unifying group is therefore $SU(3)^7$\cite{F2,F3}.
It is natural to accommodate one $SU(3)$ factor (color)
into one of the seven $SU(3)$ factors, $SU(2)_L$ as a diagonal subgroup
of two
and to identify the correctly normalized $U(1)$ 
as the diagonal subgroup of the remaining four
$SU(3)$ factors. This implies that $\alpha_2/\alpha_1 = 2$
and consequently:

\begin{equation}
\sin^2\theta = \frac{\alpha_Y}{\alpha_2 + \alpha_Y} = \frac{3/5}{2+3/5}
= \frac{3}{13} = 0.231
\end{equation}
There is a small correction for the running between $M_Z$
and the TeV scale but this is largely compensated by
the two-loop correction and
the agreement remains
as good as for SUSY-GUTS.
This is strong encouragement for the conformality approach.

\bigskip

In the second, non-abelian, example we use $\Gamma = Z_3 \times D_4$ and 
choose N=2 to arrive at a unification based on
the Pati-Salam group $SU(4)_C \times SU(2)_L \times SU(2)_R$
instead of the trinification $SU(3)^3$. This is possible
because this non-abelian $\Gamma$ has two-dimensional
representations as well as one-dimensional ones.

By the way, the dihedral group $D_4$ consists of eight rotations which leave
a square invariant: two of the
rotations are flips about two
lines which bisect the square and
the other four are
rotations through
$\pi/2, \pi, 3\pi/2$ and $2\pi$ about the perpendicular to the plane
of the square.

In this case the low energy gauge group is thus 
$SU(4)^3 \times SU(2)^{12}$. We embed $SU(3)_{color}$ in r
of the $SU(4)$ groups where r = 1 or 2 because r = 3 leads to loss of chirality.
At the same time the $SU(2)_L$ and $SU(2)_R$ are
respectively embedded in
diagonal subgroups of p and q of the twelve $SU(2)$ factors
where p + q = 12.

Since p and q are necessarily integers it is not at all obvious
{\it a priori} that
the value of $\sin^2\theta$ can be consistent with experiment.

The values of the respective couplings at the conformality/unification
scale are now:

\begin{equation}
\alpha_{2L}^{-1}(M_U) = p \alpha_U^{-1}
\end{equation} 

\begin{equation}
\alpha_{2R}^{-1}(M_U) = q \alpha_U^{-1}
\end{equation} 

\begin{equation}
\alpha_{4C}^{-1}(M_U) = 2r \alpha_U^{-1}
\end{equation} 
The hypercharge coupling is related by
\begin{equation}
\alpha_1^{-1} = \frac{2}{5} \alpha_{4C}^{-1} +
\frac{3}{5} \alpha_{2R}^{-1}
\end{equation}

Defining $y = \ln (M_U/M_Z)$ we then find the general expression
for $\sin^2 \theta_W(M_Z)$ to be:
\begin{equation}
\sin^2 \theta_W(M_Z) = \frac{p - (19/12\pi) y \alpha_U}
{p + q + \frac{4}{3} r + (11/6\pi) y \alpha_U}
\end{equation}
Here 
\begin{equation}
\alpha_S^{-1}(M_Z) = 2r \alpha_U^{-1} - \frac{7}{2 \pi} y
\end{equation}

Using these formulas and $\alpha_S(M_Z) \sim 0.12$ we
find for the natural choices (for model building) p = 4 and r = 2
that
\begin{equation}
\sin^2\theta_W(M_Z)  \simeq 0.23
\end{equation}
again in excellent agreement with experiment.

\bigskip

It is highly non-trivial that again the gauge coupling unification
works in this case which, according to the lengthy analysis in
the second paper of \cite{FK2}, is the unique accommodation
of the standard model with three chiral families
for all non-abelian $\Gamma$ with order $g \leq 31$.

\bigskip

\section{Discussion}

The successful derivation of $\sin^2\theta_W(M_Z) \simeq 0.23$
from both the abelian orbifold (based on 333-trinification)
and the non-abelian orbifold (based on 422-Pati-Salam unification)
is strong support for further investigation of the detailed phenomenology
arising from the approach.

\bigskip

More generally, the conformality provides a rigid organizing principle
which strongly constrains all couplings and parameters
in the low-energy lagrangian. Breaking of conformal symmetry
clearly needs much more study: 
so far, soft breaking has been considered mainly because
is is technically easier but spontaneous breaking would
be more satisfactory. One can even speculate that nonsupersymmetric
${\cal N}=0$ theories could have a ``flat" direction. After all, if
any such theory is finite in the UV, it likely possesses non-renormalization
theorems derivable from a symmetry different from global supersymmetry.

\bigskip

In any case, a conformal IR fixed point at the TeV scale necessitates
new particles (of spins 1, 1/2 and 0) which await discovery
in that energy regime.

\bigskip

\section{Acknowledgements}

The organizers must be thanked for creating such an excellent
Symposium. This work was supported in part by
the US Department of Energy under Grant No. DE-FG02-97ER-41036.


\bigskip



\newpage

\begin{thebibliography}{99}

\bibitem{mandelstam}  S. Mandelstam, Nucl. Phys. {\bf B213,} 149 (1983).

\bibitem{KS}  S. Kachru and E. Silverstein, Phys. Rev. Lett. {\bf 80,} 4855
(1998).

\bibitem{maldacena}  J. Maldacena, Adv. Theor. Math. Phys. {\bf 2,} 231
(1998).

\bibitem{F1}  P.H. Frampton, Phys. Rev. {\bf D60,} 041901 (1999).

\bibitem{WS}  P.H. Frampton and W. F. Shively, Phys. Lett. {\bf B454,} 49
(1999).

\bibitem{CV}  P.H. Frampton and C. Vafa, {\tt hep-th/9903226}.

\bibitem{F2}  P.H. Frampton, Phys. Rev. {\bf D60,} 085004 (1999).

\bibitem{F3}  P.H. Frampton, Phys. Rev. {\bf D60,} 121901 (1999).
\bibitem{chiral}
P.H. Frampton and S.L. Glashow, Phys. Lett. {\bf 190B,} 157 (1987);\\
Phys. Rev. Lett. {\bf 58,} 2168 (1987).
\bibitem{331}
P.H. Frampton, Phys. Rev. Lett. {\bf 69,} 2889 (1992).
\bibitem{hexagon}
P.H. Frampton and T.W. Kephart, Phys. Rev. Lett. {\bf 50,} 1343 (1983); {\it ibid} {\bf 50,} 1347 (1983).
\bibitem{guts}
Another (non-minimal) approach to family symmetry
appeared in:
P.H. Frampton, Phys. Lett. {\bf B88,} 299 (1979).
\bibitem{Pak}
S. Pakvasa and H. Sugawara, Phys. Lett. {\bf B73,} 61 (1978).
\bibitem{books}  Useful sources of information on the finite groups include:%
\newline
D.E. Littlewood, {it The Theory of Group Characters and Matrix
Representations of Groups} (Oxford 1940);\newline
M. Hamermesh, {\it Group Theory and Its Applications to Physical Problems}
(Addison-Wesley, 1962);\newline
J.S. Lomont, {\it Applications of Fimite Groups} (Academic, 1959), reprinted
by Dover (1993);\newline
A.D. Thomas and G.V. Wood, {\it Group Tables} (Shiva, 1980).
\bibitem{FK}
P.H. Frampton and T.W. Kephart, Int. J. Mod. Phys. {\bf A10,}
4689 (1995).
\bibitem{PS}  J.C. Pati and A. Salam, Phys. Rev. {\bf D10,} 275 (1974).%
\newline
R.N. Mohapatra and J.C. Pati, Phys. Rev. {\bf D11,} 566 (1975).\newline
R.N. Mohapatra and G. Senjanovic, Phys. Rev. {\bf D12,} 1502 (1975)
\bibitem{V}  H. Verlinde, Nucl. Phys. {\bf B580,} 264 (2000). 
\bibitem{RS}  L. Randall and R. Sundrum, Phys. Rev. Lett. {\bf 83} 3370 (1999);
{\it ibid} {\bf 83,} 4690 (1999).\newline
J. Lykken and L. Randall, JHEP 0006 (2000) 014. 
\bibitem{GW}
W.D. Goldberger and M. B. Wise, Phys. Rev. {\bf D60,} 107505 (1999);
Phys. Rev. Lett. {\bf 83,} 4922 (1999); Phys. Lett. {\bf B475,} 275 (2000).
\bibitem{antoniadis}
I. Antoniadis, Phys. Lett. {\bf B246} (1990) 377; \newline
I. Antoniadis and K. Benakli, Phys. Lett. {\bf B326} (1994) 69;\newline
I. Antoniadis, K. Benakli and M. Quiros, Phys. Lett. {\bf B331} (1994) 313.%
\newline
J. D. Lykken, Phys. Rev. {\bf D54} (1996) 3693.\newline
I. Antoniadis {\it et al}, Phys. Lett. {\bf B436,} 257 (1998).\newline
I. Antoniadis, S. Dimopoulos, A. Pomarol and M. Quiros, Nucl. Phys. {\bf %
B544,} 503 (1999).\newline
K. Dienes, E. Dudas and T. Gherghetta, Phys. Lett. {\bf B436} (1998) 55;
Nucl. Phys. {\bf 537,} 47 (1999); Nucl. Phys. {\bf B543,} 387 (1999).\newline
P.H. Frampton and A. Rasin, Phys. Lett. {\bf 460B,} 313 (1999).\newline
D. Ghilencea and G.G. Ross, Phys. Lett. {\bf B442} (1998) 165.\newline
C.D. Carone, Phys. Lett. {\bf B454,} 70 (1999).\\
C. Bachas, JHEP {\bf 981,} 023 (1998). \\
G. Shiu and S.H.H. Tye, Phys. Rev. {\bf D58} (1998) 106007. \newline
Z. Kakushadze and S.H.H. Tye, Nucl. Phys. {\bf B548,} 180 (1999). \newline
Z. Kakushadze, Nucl. Phys. {\bf B551,} 549 (1999).\newline
N. Arkani-Hamed, S. Dimopoulos and G. Dvali, Phys. Lett {\bf 429} (1998)506.
\bibitem{FK2}
P.H. Frampton and T.W. Kephart, Phys. Lett. {\bf B485,} 403 (2000)
and Phys. Rev. {\bf D} (2001, in press) {\tt hep-th/0011186}.
\bibitem{FMS}
P.H. Frampton, R.N. Mohapatra and S. Suh. {\it hep-ph/0104211}. submitted to Phys. Lett. {\bf B}.
\bibitem{Wil}
S. Dimopoulos, S. Raby and F. Wilczek, Phys. Rev. {\bf D24,} 1681 (1981).\\
N. Sakai, Z. Phys {\bf C11,} 153 (1981).\\
U. Amaldi, W. De Boer and H. Furstenau, Phys. Lett. {\bf B260,} 447 (1991).\\
U. Amaldi, W. De Boer, P.H. Frampton, H. Furstenau and J.T. Liu, Phys. Lett. {\bf B281,} 374 (1992).
\end{thebibliography}
\end{document}